\documentclass[a4paper,twocolumn,11pt,accepted=2024-10-10]{quantumarticle}
\pdfoutput=1
\usepackage{amsmath}
\usepackage{amssymb}
\usepackage{graphicx}
\usepackage{dcolumn}
\usepackage{bm}
\usepackage{xcolor}
\usepackage{tikz}
\usepackage{xfrac}
\usepackage{blindtext}

\definecolor{darkblue}{HTML}{004D6B}
\definecolor{darkred}{HTML}{8c1515}

\usepackage{hyperref}
\hypersetup{
	colorlinks=true,
	urlcolor=darkred,
	citecolor=darkblue,
	linkcolor=darkred,
	breaklinks
}

\usepackage{physics}
\usepackage{bbold} 

\usepackage[numbers]{natbib}

\begin{document}

\title{Programmable adiabatic demagnetization for systems with trivial and topological excitations}

\author{Anne Matthies}
\affiliation{Institute for Theoretical Physics, University of Cologne, 50937 Cologne, Germany}
\affiliation{Department of Condensed Matter Physics, Weizmann Institute of Science, Rehovot, 76100, Israel}
\orcid{0000-0003-1155-1281}
\author{Mark Rudner}

\affiliation{Department of Physics, University of Washington, Seattle, WA 98195-1560, USA }
\orcid{0000-0002-5150-6234}
\author{Achim Rosch}
\affiliation{Institute for Theoretical Physics, University of Cologne, 50937 Cologne, Germany}
\orcid{0000-0002-6586-5721}
\author{Erez Berg}
\affiliation{Department of Condensed Matter Physics, Weizmann Institute of Science, Rehovot, 76100, Israel}
\orcid{0000-0001-8956-3384}

\begin{abstract}
	We propose a simple, robust protocol to prepare a low-energy state of an arbitrary Hamiltonian on a quantum computer or programmable quantum simulator. 
	The protocol is inspired by the adiabatic demagnetization technique, used to cool solid-state systems to extremely low temperatures. A fraction of the qubits (or spins) is used to model a spin bath that is coupled to the system.  
	By an adiabatic ramp down of a simulated Zeeman field acting on the bath spins, energy and entropy are extracted from the system. The bath spins are then measured and reset to the polarized state, and the process is repeated until convergence to a low-energy steady state is achieved. We demonstrate the protocol via  application to the quantum Ising model. We study the protocol's performance in the presence of noise and show how the information from the measurement of the bath spins can be used to monitor the cooling process. The performance of the algorithm depends on the nature of the excitations of the system; systems with non-local (topological) excitations are more difficult to cool than those with local excitations. We explore the possible mitigation of this problem by trapping topological excitations.
\end{abstract}
\maketitle


Ground state preparation on quantum simulators and computers is very important for the characterization of ground state properties in quantum chemistry and material science~\cite{Lloyd1996,Reiher2017,Cao2019,McArdle2020}. 
More generally, it can also be utilized for a variety of quantum information problems~\cite{Farhi2014,Biamonte2017,Montanaro2016}. On a quantum computer, the Hamiltonian can be digitally implemented using a sequence of unitary gates. Many approaches of ground state preparation have been proposed, including variational quantum simulation \cite{Peruzzo2014,McClean2016,Cerezo2021,Tilly2022,Fedorov2022} and adiabatic state preparation \cite{Nishimori98,Farhi2000,Childs2001,AspuruGuzik2005,Albash2018}. Each of these approaches has its own challenges. For example, the performance of variational quantum simulation highly depends on the quality of the variational ansatz \cite{Lee2022,Fedorov2022}. At the same time, adiabatic state preparation is sensitive to the trajectory in Hamiltonian space connecting the initial and final Hamiltonians. It tends to fail when a phase transition separates the initial and final states. Furthermore, errors that occur during adiabatic state preparation are not intrinsically corrected and may be difficult to detect.

 Recently, algorithms that mimic cooling by coupling to a simulated low-entropy bath~\cite{Boykin2002,Kaplan2017,Wang2017,Feng2022,Polla2021,Zaletel2021,Metcalf2020,RodriguezThesis} 
 have been proposed as alternative routes that may overcome some of these challenges. The key advantages of such cooling algorithms are that they can be run cyclically without any special requirements on initial states and do not require prior knowledge of the target ground state. 
 Simulated cooling schemes also do not require the target state to be adiabatically connected to a product state. By construction, cooling schemes target the ground state manifold but cannot be used directly to prepare one specific state within a degenerate ground-state manifold. The latter is, however, rarely required in the context of quantum simulations. More important is another property of the algorithm.
 The cyclic operation provides inherent robustness to weak noise by automatically correcting some errors and removing unwanted excitations on subsequent cycles. Furthermore, in simulated cooling on a quantum computer, monitoring the bath spins during the cooling protocol can provide information on the process of approaching the system's ground state without observing the system directly and may be used to identify the presence of errors.
\begin{figure}[ht]
	\begin{minipage}{0.97\columnwidth}
		\includegraphics[width = \textwidth]{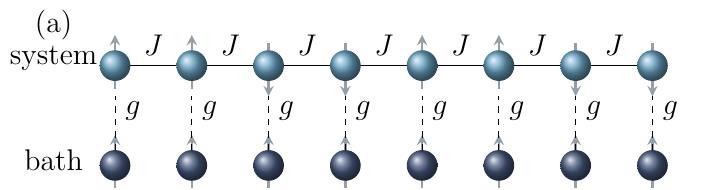}
	\end{minipage}
 \hfill
	\begin{minipage}{0.97\columnwidth}
 \centering
		\includegraphics[width = \textwidth]{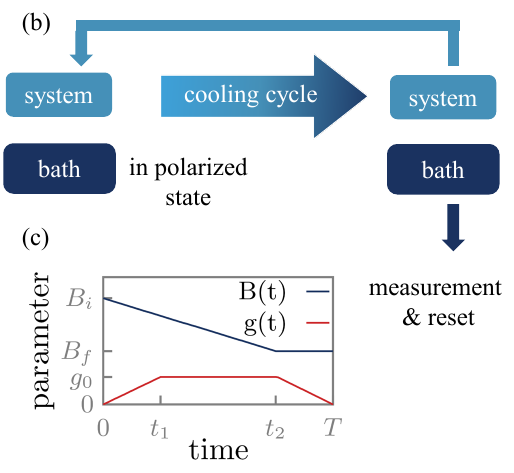}
	\end{minipage}
 	\caption{
Schematic illustration of the proposed setup. (a) Each system spin $s_i$ (light blue) is coupled to a bath spin $\sigma_i$ (dark blue) by switching on the coupling $g(t)$. System spins may be generically coupled, 
as illustrated here by $J$. At the beginning of the protocol, the bath is polarized and subjected to a large simulated magnetic field $B(t)$, 
while the system may begin in a random state. Panel (b) illustrates the cooling cycle, and panel (c) shoes the time dependence of the parameters. A magnetic field $B(t)$ is applied to  the bath spins, and linearly decreased from $B_0$ to $B_f$ until time $t_2=3T/4$, after which it is held constant until time $T$, the total duration of the sweep. 
The system-bath coupling $g(t)$ is slowly switch on until $t_1=T/4$, kept constant at $g(t)=g_0$ until time $t_2$, and then switched off.
At the end of each sweep, the bath spins are measured and reset.}
	\label{ground state:intro}
\end{figure}
Here, we define a simple, scalable protocol for low-energy state preparation of an arbitrary gapped Hamiltonian on a quantum computer or programmable quantum simulator. We draw an analogy between our protocol and adiabatic demagnetization \cite{Debye1926,Giauque1927}, known from solid-state systems. For example, with nuclear adiabatic demagnetization, one can reach temperatures in the $\mu$K range \cite{Cao2021, Jones2020} by coupling the system to nuclear moments polarized in a large magnetic field and adiabatically ramping down the field. 

Fig.~\ref{ground state:intro} illustrates the cooling scheme.
We consider a generic system comprised of $N$ qubits (light blue), coupled to an additional set of $N_{\rm bath}$ ``bath'' qubits (dark blue).
We take the number of bath qubits to be a fixed fraction of the number of system qubits; for simplicity, we assume $N_{\rm bath} = N$ throughout (i.e., one bath qubit per system qubit, as shown in Fig.~\ref{ground state:intro}). At the beginning of each cooling cycle, the bath spins are prepared in a fully polarized product state and subjected to a simulated Zeeman field along the polarization axis. By an adiabatic ramp down of the magnetic field, we change the energy splitting of the bath spins and activate processes that transfer energy and entropy from the system to the bath spins. After each cycle, the bath is adiabatically decoupled from the system and then reset to the polarized state.
Measuring the state of the bath spins before resetting them provides information about the progress of the cooling protocol without affecting the state of the system spins and can be used to monitor the process and detect errors.

Recently, similar cooling protocols have been studied for the cases of a single bath spin~\cite{Polla2021}, and many bath spins with non-local all-to-all couplings with the system qubits~\cite{Kaplan2017}.
By considering the number of bath spins to grow as a constant fraction of the system size, with simple to-implement local couplings, we obtain a scalable approach with a cooling rate independent of system size. 

We present a perturbative analysis characterizing the efficiency of our cooling scheme in the linear response regime, and numerically demonstrate its effectiveness as well as the steady states that result from cyclic operation by applying it to the one-dimensional quantum Ising model. 
This model hosts a paramagnetic and a ferromagnetic phase, and can be tuned to be integrable or self-thermalizing. 

At suitably chosen (dual) points in the phase diagram, the paramagnetic and ferromagnetic phases host identical excitation spectra, but the natures of their excitations are qualitatively different. The low-energy excitations of the paramagnet 
can be approximately viewed as spin flips, which can be created and destroyed locally. 
In contrast, the elementary excitations of the ferromagnet are domain walls which are non-local and topological. In the bulk, domain walls can only be created and destroyed in pairs. 
We use these features to illustrate how the presence of 
topological excitations affects the efficiency of state preparation.
Importantly, we also characterize the robustness of the steady state to noise and imperfect operations within the field sweeps.

 Preparing topological ground states with long-ranged entanglement on a quantum simulator is especially challenging because single topological bulk excitations cannot be removed locally. Thus, any cooling algorithm that relies on a local coupling of the system and bath spins is expected to be parametrically slower for systems with topological excitations than those with topologically trivial (i.e., local) excitations.
 At the same time, the presence of mobile topological excitations is highly detrimental to quantum memories and quantum simulations employing the topological nature of the system. Here, we propose and numerically demonstrate the possibility of trapping topological excitations as a way to mitigate this problem.

\emph{Protocol --} Our central goal is to prepare the ground state or a low-energy state of a many-body quantum system with Hamiltonian $H_s$ realized on a programmable quantum simulator. Local operators $\hat{A}^s_i$ 
of the system are coupled to the bath qubits described by the spin operators $\boldsymbol \sigma_i$,
\begin{align}
	H=H_s + \sum_{i=1}^N \Large[g(t) \hat{A}^s_i \sigma_i^y -B(t)\sigma_i^z \Large],
 \label{eq:H}
\end{align}
where $g(t)$ is a time-dependent system-bath coupling and $B(t)$ is the time-dependent simulated Zeeman field  that acts on the bath spins.

 Initially, the system is in an arbitrary state, and the bath is prepared in the fully polarized state, $|\uparrow\uparrow\uparrow\dots\rangle$. The cooling protocol consists of $N_c$ 
identical cycles, where the bath effective Zeeman field $B(t)$ and system-bath coupling $g(t)$ are swept according to the schedule shown in Fig.~\ref{ground state:intro}b.
The system-bath coupling $g(t)$ is increased linearly for $0<t<t_1$, held constant for $t_1<t<t_2$, and then decreased linearly back to zero for $t_2<t<T$. 
Simultaneously, the magnetic field is linearly decreased from $B_i$ to $B_f$. At the end, the bath spins are measured and reset to the polarized state, and a new cycle starts. 

For ideal preparation (see below for a discussion of noise/errors), the bath begins each sweep in a state with zero entropy. The gradual reduction of the magnetic field acting on the bath spins, in the presence of the system-bath coupling, causes energy and entropy to be transferred from the system into the bath.
The measurement and reset of the decoupled bath at the end of the cycle remove the transferred entropy and allow to restart the cycle with a zero entropy bath. Furthermore, information from the measurements can be used to monitor and improve the cooling process, see below.

\emph{Pertubative analysis.--} 
Assuming a small system-bath coupling $g$, we derive an analytical expression for the cooling rate  $\Gamma_c(t)=-\frac{d}{dt}\langle{H_{s}}\rangle=- \langle i [H_s, H]\rangle$. Within linear response, the cooling rate at time $t$ is obtained as \cite{Suppl}
\begin{widetext}
\begin{align}\label{eq:generalIntegral}
	\Gamma_c= -ig(t)\!\int_{0}^t \!\! \! dt'\, g(t')\sum_{i} \langle \psi|[\dot{\hat{A}}^s_i(t) \sigma_i^y(t),\hat{A}^s_i(t')\sigma_i^y(t')] |\psi \rangle,
\end{align}
\end{widetext}
with $\dot{\hat{A}}^s_i(t)=i[H_s,\hat{A}^s_i(t)]$. All time-dependent operators are evaluated in an interaction picture using $H(t)$ at $g=0$ as the generator of dynamics. In Eq.~(\ref{eq:generalIntegral}), $|\psi\rangle$ is the state of the system at the beginning of the cooling cycle, where system and polarized bath spins decouple. 
If $H_s$ is a generic (non-integrable) interacting many-particle system with an equilibration time short compared to $T$, the initial state can approximately be described by a thermal state with a temperature that is reduced in each step of the cooling cycle.

In the supplementary material \cite{Suppl}, we analyze the integral in Eq.~(\ref{eq:generalIntegral}) for a time-dependent $B(t)$. For slow changes of $g(t)$ and $B(t)$, one can use a stationary phase approximation to perform the integral, where the result for self-equilibrating systems simplifies to 
\begin{align}
&\frac{\Gamma_c}{N} \approx \label{eq:PT} \\ &- 4\pi g(t)^2\!  \int \!  d\omega  [1+n_B(\omega)]\omega \chi^{\prime\prime}_s(\omega) \delta\big(\omega+2B(t)\big),\nonumber
\end{align}
 where $n_B(\omega)$ is the Bose function and  $\chi^{\prime\prime}_s(\omega)=\frac{1}{N} \sum \chi^{\prime\prime}_{s,i}(\omega)$ is the imaginary part of the local, retarded $\hat{A}^s_i$ susceptibility averaged over $i$, with $\chi_{s,i}(t)=i \theta(t) \langle [\hat{A}^s_i(t),\hat{A}^s_i(0)] \rangle$. Both $n_B(\omega)$ and $\chi_{s,i}$ are evaluated for a thermal state at the effective temperature of the system.
 
  Equation \eqref{eq:PT} shows that energy is mainly extracted via resonant processes, which occur when the bath spins' energy splitting $2 B(t)$ matches an excitation energy in the system.
Thus, as $B(t)$ is ramped down during each sweep, high-energy and then lower-energy excitations are removed from the system.
The total energy extracted in one sweep, $\int_0^T dt\, \Gamma_c$, can be evaluated by replacing the time integral with an integral over the ramping field $B(t)$, $\int dt = \int \frac{d B}{\dot B}$. Thus, the extracted energy is
 is  proportional to $g^2/{\dot B}\propto g^2 T$, for large sweep durations $T$.  
 In deriving Eq.~(\ref{eq:PT}), we assumed that the sweep time $T$ is long compared with the self-thermalization time of the system and that the coupling is weak such that less than one spin flip occurs per spin per field sweep. In the supplementary material \cite{Suppl}, we furthermore show that one can efficiently extract energy even for relatively fast sweep rates (or, equivalently, short time scales $T$) set by the bandwidth and gap of the system.
 
While the perturbative analysis given above relies on the small-$g$ limit, the cooling protocol itself does not require $g(t)$ to be small. 
The only requirement is that the time dependence of the Hamiltonian is adiabatic (in the thermodynamic sense). To improve the cooling rate, it is therefore useful to choose $g$ as large as possible while staying in the adiabatic regime (keeping the overall time duration $T$ fixed). Thus, one should also avoid regimes where the coupling $g$ induces a phase transition in the coupled system-bath Hamiltonian,  \eqref{eq:H}.

\emph{Numerical analysis:} We exemplify our protocol via an application to the one-dimensional quantum Ising model: 
\begin{align}
	H_s=-\sum_{i} J s_i^z s_{i+1}^z - \sum_{i} \left(h_x s_i^x + h_z s_i^z \right),     \label{eq:Ising}
\end{align}
with an exchange coupling $J$, transverse field $h_x$ and longitudinal field $h_z$.  The latter can be used to make the  model non-integrable. We use periodic boundary conditions unless specified otherwise. Each bath spin $\bm{\sigma}_i$ is coupled to the local degree of freedom of the system $\hat{A}_i^s=s_i^y$.

To simulate the cooling protocol in the presence of noise, we use the stochastic Schrödinger equation.  The time-evolution operator is approximated using the 2nd order Suzuki-Trotter decomposition $U=\prod_n U_n$, where $n$ labels the Trotter time step (of duration $\Delta \tau$), and 
\begin{align}
	U_n&=\\ &e^{-i \frac{\Delta\tau}{2} H_Z(t_n)} e^{-i \Delta\tau H_Y (t_n)}e^{-i \Delta\tau H_X(t_n)}e^{-i \frac{\Delta\tau}{2} H_Z(t_n)}.\nonumber
	\end{align}
Here $H_{X,Y,Z}$ are the terms in the Hamiltonian that contain the $X$, $Y$, or $Z$ components of both the system spins and the bath spins. For each spin, we randomly apply one of the Pauli operators after the first unitary $U(H_Z)$ and the third unitary $U(H_Y)$ of each Trotter step with probability $p_{\text{err}}$, thus realizing depolarizing noise. For our implementation, we choose $\Delta \tau=0.06$ to approximate a continuous time evolution, which can be realized experimentally, e.g., on a quantum simulator based on Rydberg atoms. For implementation on a digital quantum computer, one would instead choose larger  $\Delta \tau$ to reduce the number of gate operations.

After the end of a sweep, consisting of $N_\tau$ Suzuki-Trotter steps, a projective measurement of all bath spins is performed, leaving the system in a pure state.
The bath spins are then reset back to the fully polarized state. The measurement results are useful for monitoring the cooling progress, see below. 

\begin{figure*}[t]
\centering
	
	\includegraphics[width=1.47\columnwidth]{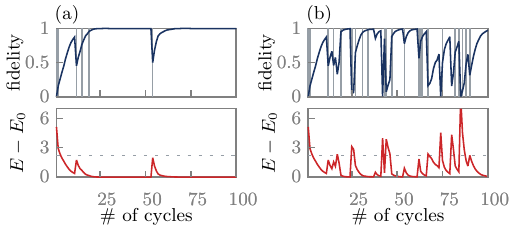}

	\caption{\label{fig:single run}Single run of the protocol for the non-integrable ($J=1$, $h_x=1$, $h_z=0.2$, $N_s=8$, $N_{\tau}=101$, $T=6$, $B_i=5$, $B_f=0.7$, $g_0=0.5$) case up to 100 protocol cycles. (a) No external noise is applied, i.e., only adiabatic and Trotter errors occur. (b) External noise of $2\cdot 10^{-2}$ per sweep and spin is applied. The top panels show the system ground state fidelity (blue). Vertical grey lines denote cycles when a flipped spin is observed in the bath at the end of a cycle.
	The bottom panels show the difference between the expectation value of the system's energy at the end of each cycle and its ground state energy (red). The energy gap of the system is indicated by the dashed grey line.   
    At the noise level used in panel (b), the system energy remains below the gap, on average. 
	}
\end{figure*}

In Fig.~\ref{fig:single run}, we show the expectation value of the system energy relative to its ground state energy value $E_0$, $E-E_0$, and the ground state fidelity of the system at the end of each cycle for a single trajectory, as a function of the cycle index, both (a) without and (b) with noise. 
In the upper panel, we display a vertical gray line to mark the cycles where, after measuring the bath spins, at least one spin is found to have flipped.
 The system and bath include $N=8$ spins each.
 All other parameters are specified in the caption of Fig.~\ref{fig:single run}.

 Without noise ($p_{\rm{err}}=0$),  Fig.~\ref{fig:single run}a, 
 $E-E_0$
 drops below the energy gap of the system (horizontal dashed line) already after 3-4 cycles. 
 The fidelity takes about $20$ cycles to saturate to a value close to 1. Occasionally, the fidelity suddenly decreases due to the combination of Trotter and/or adiabatic errors and the inherent randomness coming from the measurement of the bath spins. 
 Each of these jumps is accompanied by at least one flipped bath spin (vertical lines); the observation of flipped bath spins thus heralds the presence of errors and can be used to increase the fidelity of the prepared state via (post) selection.

The outcomes of the bath spin measurements can be used 
both as an effective thermometer for the system and as a stopping criterion of the protocol. The bath spins should remain fully polarized at the end of each sweep once the system has reached the ground state for a perfect adiabatic process and in an ideal setting without errors.
Therefore, if the bath spins remain fully polarized over several consecutive cycles, we can conclude that the system's ground state has been reached with high confidence. 
In the presence of noise, we can still enhance the protocol's performance by using a similar criterion.
 
To demonstrate the protocol's robustness to errors, in Fig.~\ref{fig:single run}b we show results for a trajectory with an imposed error rate of $p_{\text{err}}=10^{-4}$ per spin (corresponding to an average number 
 $\eta_e=2\cdot 10^{-2}$ of errors per sweep per spin). Although jumps of energy and ground state fidelity occur more frequently than in the case without errors, the cooling protocol is still able to recover after noise events, with the average number of excitations remaining below $1$ in the steady state.

\begin{figure*}[]
	\centering{
	\includegraphics[width=1.77\columnwidth]{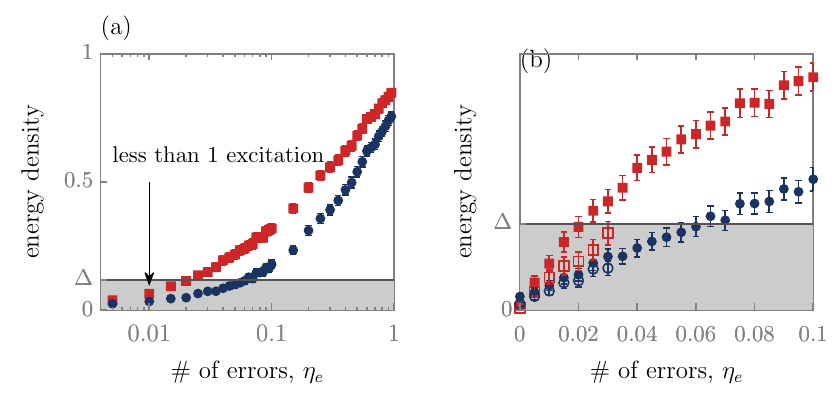}}
	\caption{\label{fig:energy noise}$(E-E_0)/E_0$ as a function of the average number of errors $\eta_e$: (a) The ferromagnetic case ($J=1$, $h_x=0.5$, $h_z=0$) is shown in red squares, the paramagnetic case ($J=0.5$, $h_x=1$, $h_z=0$) in blue dots, and the non integrable case ($J=0.4$, $h_x=0.5$, $h_z=0.8$) in orange triangles. Within the grey shaded area, the energy is below the lowest excitation gap $\Delta$.  
	(b) For small noise levels, post-selection (empty red squares: FM, empty blue circles: PM) improves the outcome of the protocol, i.e., it lowers the energy of the system. See text for the description of the post-selection protocol. (Parameters: $N_s=8$, $N_{\tau}=101$, $N_c=100$, $N_{init}=1000$, $T=6$, $B_i=5$, $B_f=0.7$, $g_0=0.5$).
	}
\end{figure*}
\twocolumn
To explore how our cooling protocol performs on systems with and without topological excitations (domain walls), in Fig.~\ref{fig:energy noise}, we plot the energy density $e=(E-E_0)/E_0$ averaged over $N_{init}=1000$ trajectories, as a function of the noise rate in the ferromagnetic (red squares) and paramagnetic (blue circles) regimes. For small noise levels, $e$ in the steady state is linear in the noise rate with a slight offset due to finite sweep rates and Trotter errors. As shown in the zoom-in in panel (b), for noise levels below 0.02 and 0.05 for the ferro- and paramagnetic cases, respectively, less than one excitation remains in our 8-site system. We also show that in this regime, $e$ can be further reduced by about 20\%-50\% by post-selection (empty symbols): the cooling protocol was stopped if, in 5 consecutive sweeps, no spin-flip of bath spins was measured. 

While our numerical simulations are limited to small systems, we can obtain a qualitative characterization of the performance of cooling protocols in the presence of noise from a simple rate equation for the density of excitations $n=N_{\rm{ex}}/V$ (where $N_{\rm{ex}}$ is the number of excitations, and $V$ is the system's volume), 
\begin{align}
    \partial_t n = \Gamma_{\text{noise}}-\gamma_c n^M. \label{rateEq}
\end{align}
The first term on the right-hand side describes the creation of excitations 
with a rate $\Gamma_{\rm noise}$, which is linear in the error rate. The second term encodes cooling via a mechanism where $M$ adjacent point-like excitations are removed simultaneously.
Crucially, while certain types of local excitations can be removed one at a time ($M=1$), topological excitations can only be removed as pairs or higher-order clusters ($M>1$). We note that values of $M>1$ may also characterize topologically-trivial excitations: for example, the annihilation of particle and hole excitations in a semiconductor is an $M=2$ process, while bound excitons can be removed individually, $M=1$.
In the steady state, the excitation density is thus given by \begin{align}
    n=\left(\Gamma_{\text{noise}}/\gamma_c\right)^{1/M}. \label{nnoise}
\end{align}

Cooling is most effective when single excitations can be removed by the coupling to the bath, i.e., $M=1$. $M=1$ is the case in the paramagnetic phase. 
In contrast, the excitations in the ferromagnetic phase are domain walls, which are non-local objects and can only be removed in pairs, i.e., $M=2$.  

\begin{figure*}[t]
	\centering
	\includegraphics[width=1.47\columnwidth]{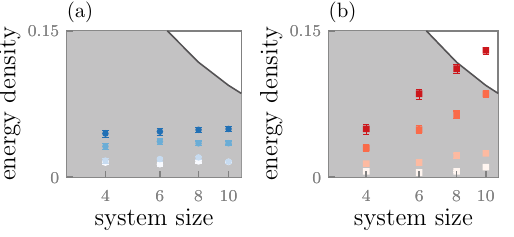}

	\caption{\label{fig:energy systemsize}
	Energy density  $(E-E_0)/E_0$ as a function system size for (a) the paramagnetic case and (b) the ferromagnetic case and for an average number of errors $\eta_e=0$, $2\cdot 10^{-3}$, $10^{-2}$ and $2\cdot10^{-2}$ per sweep and spin (light to dark). In the shaded grey area the average energy is below the system's gap. In the paramagnetic phase, the energy density grows with $\eta_e$ but not with system size, whereas in the ferromagnetic phase the energy density grows approximately linearly with system size. The slope of the energy density vs.~system size increases with increasing $\eta_e$. 
    The observed behavior is consistent with the prediction of the rate equation model, Eqs.~(\ref{rateEq}) and (\ref{nn}).
	Parameters are the same as in Fig.~\ref{fig:energy noise}. }
\end{figure*}

The model in Eq.~(\ref{rateEq}) describes the cooling process in the thermodynamic limit. In a finite size, $d$-dimensional system with very small $n$, the probability that $M$ excitations are close to each other is proportional to $1/V^{M-1}$, where $V$ is the volume of the system. Thus in the limit $\Gamma_{\rm noise}\to 0$, when $n \lesssim M/V$, we expect
\begin{align}
n \sim \frac{\Gamma_{\text{noise}}}{\gamma_c} V^{M-1}. \label{nn}
\end{align}
We test this within our model by comparing it with results for the transverse field Ising model at dual points in the ferromagnetic and paramagnetic phases. The elementary excitations of the ferromagnet are domain walls which, for periodic boundary conditions, can only be created and annihilated in pairs, $M=2$, while $M=1$ for the paramagnet. Therefore, we choose a parameter set where the numerical values of $J$ and $h_x$ are exchanged. Using the self-duality of the transverse field Ising model ensures that the dispersion of the elementary excitations is identical in the two cases. 
Using the heuristic that in a gapped system the energy density is roughly linear in the number of excitations, we estimate $n$ by $e$.
Fig.~\ref{fig:energy noise}b  demonstrates that the energy density $e$ is linear in the noise rate in the limit of small noise for both cases. In Fig.~\ref{fig:energy systemsize}, we show that the dependence of $e$ on the system size is consistent with Eq.~(\ref{nn}): the excitation density in the paramagnetic phase $(M=1)$ is nearly independent of system size, while for the ferromagnetic phase $(M =2)$, $e$ grows linearly with system size. 
While in Fig.~\ref{fig:energy noise}  results for two dual integrable model are shown, we find that the cooling protocol shows a similar performance for non-integrable models, see supplementary material \cite{Suppl}.

Our results highlight the fact that it is generally difficult to remove topological excitations.  
We now discuss a way to partially mitigate this problem. 
Instead of removing the topological excitations, they can be localized in specially designed ``traps,'' thus reducing the number of free (mobile) excitations. For trapping, one has to transfer a mobile high-energy excitation to a low-energy state at a trapping site. As the number of topological excitations is not changed in this process, this can be accomplished by our cooling protocol, transferring the excess energy to the bath spins.
\begin{figure*}[t]
	\centering
	\includegraphics[width=1.47\columnwidth]{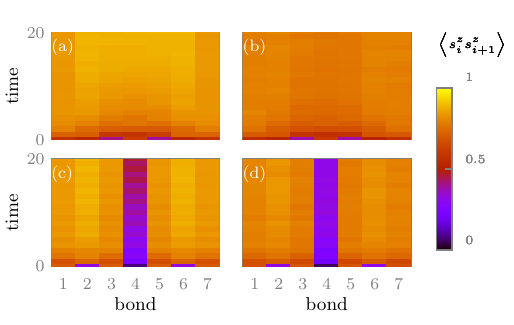}

	\caption{\label{fig:trap} Bond correlation function $\left<s_i^zs_{i+1}^z\right>$ for the ferromagnetic case ($J=1$, $h_x=0.5$, $h_z=0$) with open boundary conditions with no noise (left) and an average number of errors of $\eta_e=2\cdot 10^{-2}$ per sweep and spin (right). There is no trap for data in panel (a) and (b). For panel (c) and (d) the trap is implemented by an decreased $J$ coupling $J_{trap}=0.33J$ between the $4$th and $5$th spin. Parameters: $N_s=8$, $N_{\tau}=101$, $N_c=100$, $N_{init}=1000$, $T=6$, $B_i=5$, $B_f=0.7$, $g_0=0.5$.
	}
\end{figure*}

In the ferromagnetic case, a domain wall can be bound to a specific bond by setting the exchange coupling on that bond to $J_{\text{trap}} < J$. In Fig.~\ref{fig:trap}, we show the effect of such a trap on a system with open boundary conditions. The left panels show the expectation value $\langle s_i^z s_{i+1}^z\rangle$ for each bond at the end of each cycle. 
Initially, the system is in a highly excited state.  
After a few cycles, $\langle s_i^z s_{i+1}^z\rangle \approx 1$ on all bonds. 
If there is no trap, the value of $\langle s_i^z s_{i+1}^z\rangle$ is reduced for the noisy system [Fig.~\ref{fig:trap}(b)] compared with its value for the noiseless system [Fig.~\ref{fig:trap}(a)]. The smooth profile of $\langle s_i^z s_{i+1}^z\rangle$ in space indicates that domain walls are delocalized across the chain. As shown in Figs.~\ref{fig:trap}(c) and (d),  when we create a trap in the middle of the chain, a domain wall tends to be localized at that spot. While creating a trap changes the Hamiltonian and locally lowers the gap, the localization in trapping sites can be a very efficient methods to remove mobile excitations.

\begin{figure*}[t]
	\centering
		\includegraphics[width=1.47\columnwidth]{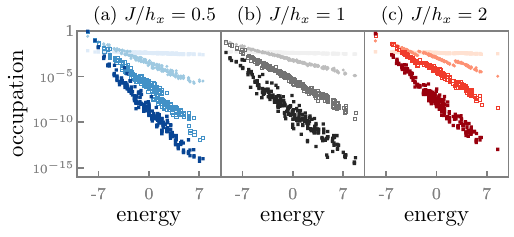}

	\caption{\label{fig:energy occupation} Occupation of system energy eigenstates for noise rates $\eta_e=0$, $0.02$, $0.2$, $2$ per sweep and spin (dark to light) for (a) the paramagnetic {($J=0.5$, $h_x=1$, $h_z=0$)}, (b) the non-integrable {($J=1$, $h_x=1$, $h_z=0.2$)} and (c) the ferromagnetic case {($J=1$, $h_x=0.5$, $h_z=0$)}. The distribution is obtained by averaging over noise configurations and measurement outcomes.
	{Parameters: $N_s=8$, $N_{\tau}=101$, $N_{init}=1000$, $T=6$, $B_i=5$, $B_f=0.7$, $g_0=0.5$. }}

\end{figure*}

To further characterize the steady states reached by our protocol, in Fig.~\ref{fig:energy occupation}, we plot the populations of the system's energy levels for several values of the noise rate. Here, periodic boundary conditions are used.  
While our protocol produces non-equilibrium states, in the paramagnetic phase, we observe a roughly thermal distribution (i.e., the probability decreases exponentially with energy), even when the unperturbed system is integrable. 
In the ferromagnetic phase, however, there is a striking deviation from a thermal distribution: Fig.~\ref{fig:energy occupation}c shows jumps in the distribution function. The difficulty of eliminating domain walls can explain these jumps: the system equilibrates in distinct sectors with 0, 2, and 4 domain walls present.

\emph{Conclusion:} We have presented a scalable protocol to prepare low-energy states of an arbitrary Hamiltonian on a quantum computer or programmable analog quantum simulator. 
The protocol produces a steady state that approaches the ground state in the low noise limit, assuming that adiabatic and Trotter errors can be controlled. 
The protocol involves measuring and resetting the bath spins at each cycle.
This non-unitary element of the cooling protocol is crucial as it effectively extracts entropy, disentangles bath and system, and allows to repeat the cycle again and again. While the algorithm also works if one replaces the combination of measurement and reset by only a reset  of bath spins, one can use the measurement results to gain information about the state of the system without collapsing it. Cooling is successful when fewer and fewer bath spins are flipped on successive cycles. 
Furthermore, by conditioning the termination of the protocol on measurement outcomes, one can improve the fidelity of the prepared state.

Systems with topological excitations are inherently more difficult to cool, as such excitations can only be removed when more than one excitation is present near a given position. 
Interestingly, this property offers a possibility to identify the presence of topological order within our protocol. In general, topological order is notoriously difficult to detect.
Within our algorithm, one can measure the efficiency of the cooling process; if the number of excitations is non-linear in the noise level [$M\ge 2$ in Eq.~\eqref{nnoise}], this indicates that excitations cannot be created and destroyed locally. Thus, one may use measurements of ``coolability'' in a quantum simulation experiment to detect topological phases.

It is interesting to compare the performance of our programmable cooling algorithm to adiabatic state preparation \cite{Nishimori98,Farhi2000,Childs2001,AspuruGuzik2005,Albash2018} in the presence of a finite noise rate $\Gamma_{\text{noise}}$. 
Here we assume that the ground state of the target Hamiltonian $H_t$ is separated by a quantum phase transition from the initial state, assumed to be the groundstate of a Hamiltonian $H_i$. In the presence of noise, one has to optimize the duration $T$ of the adiabatic state preparation protocol, where one slowly changes the Hamiltonian from $H_i$ to $H_t$.
For longer $T$, less excitations are created via the Kibble-Zurek mechanism \cite{Kibble1976,Zurek1985}
when crossing the phase transition, but more defects arise due to the finite noise rate, an effect which has previously be called anti-Kibble-Zurek behavior \cite{Griffin2012,Dutta2016}. Optimizing for $T$ and using the Kibble-Zurek formula applied to quantum phase transitions \cite{Dziarmaga2005,Zurek2005,Polkovnikov2005}, one obtains \cite{Dutta2016} that point-like defects occur at the density
\begin{align}
    n \sim \min_T\left[\Gamma_{\text{noise}} T + \frac{c}{T^{\frac{d \nu}{1+z \nu}}} \right]\sim \Gamma_{\text{noise}}^{\frac{d \nu}{1 + (d + z) \nu}}
\end{align}
where $\nu$ and $z$ are the correlation length and dynamical critical exponent of the quantum critical point in spatial dimension $d$ and $c$ some non-universal constant.
Comparing this to programmable cooling, Eq.~\eqref{nnoise}, we find that when preparing a trivial phase, $M=1$, our cooling algorithm will {\em always} outperform adiabatic state preparation in the small $\Gamma_{\text{noise}}$ limit.
For the preparation of a topological phase where defects can only be annihilated in pairs,
the adiabatic state preparation is more efficient than cooling only for 
\begin{align}
    \nu >\frac{1}{d-z}, \qquad M=2.
\end{align}
For state preparation in low-dimensional systems, $d=1$ or $d=2$, programmable cooling appears to be generically the better option. For example, $\nu \approx 0.63$ is much smaller than $1/(d-z)$ for quantum critical points in the Ising universality class in $d=2$ where  $z=1$. In three dimensions, $\nu=1/2$, $z=1$, the formulas given above cannot decide whether cooling or adiabatic state preparation are more efficient as they both predict $n\sim \Gamma_{\text{noise}}^{1/2}$, possibly with logarithmic corrections.

The arguments given above do not take into account that both the cooling algorithm and adiabatic state preparation can be further optimized \cite{Barankov2008,Nishimori21}. For example, {\em if} the location of the phase transition is known, one can slow down the adiabatic state preparation protocol close to the critical point in such a way that defects are created with a density proportional to $1/T^{d/z}$ \cite{Barankov2008}. After optimization with respect to $T$, this results in a defect density $n \sim \Gamma_{\text{noise}}^{d/(d+z)}$. While this is still worse than the cooling protocol of trivial states, $n\sim \Gamma_{\text{noise}}$ for $M=1$, it may outperform in $d>1$ a cooling scheme for topological states, $n\sim \Gamma_{\text{noise}}^{1/2}$ for $M=2$.

Due to its relative robustness to errors, our cooling protocol can be implemented, tested, and optimized directly on present-day quantum hardware consisting of only a few qubits. Therefore, we expect that our cooling protocol will prove useful in future quantum computers which can simulate complex Hamiltonians with unknown ground states and excitations.

\emph{Data availability:}
We note that data underlying the figures presented in this work can be retrieved at the zenodo \cite{Matthies2024}.

\emph{Acknowledgements:}
We thank Ehud Altman, Imane El Achchi, Snir Gazit, Elliot Kapit, Hidetoshi Nishimori, Vadim Oganesyan, Daniel Alcalde Puente, and Gil Refael for useful discussions. 
We acknowledge funding from the German Research
Foundation (DFG) through CRC 183 (project number
277101999, A01) and – under Germany’s Excellence Strategy – by the
Cluster of Excellence Matter and Light for Quantum
Computing (ML4Q) EXC2004/1 390534769. E.B. acknowledges support from the Israel Science Foundation Quantum Science and Technology (grant no. 2074/19) and from a research grant from Irving and Cherna Moskowitz. 
M. R. is grateful to the University of Washington College of Arts
and Sciences and the Kenneth K. Young Memorial Professorship for support. 
We furthermore thank the Regional Computing Center of the University of Cologne (RRZK) for providing computing time on the DFG-funded (Funding number: INST 216/512/1FUGG) High Performance Computing (HPC) system CHEOPS.

\bibliographystyle{quantum}
\bibliography{literature_qcByCooling.bib}

\begin{thebibliography}{10}

\bibitem{Lloyd1996}
Seth Lloyd.
\newblock ``Universal quantum simulators''.
\newblock \href{https://dx.doi.org/10.1126/science.273.5278.1073}{Science {\bf
  273}, 1073--1078}~(1996).

\bibitem{Reiher2017}
Markus Reiher, Nathan Wiebe, Krysta~M. Svore, Dave Wecker, and Matthias Troyer.
\newblock ``Elucidating reaction mechanisms on quantum computers''.
\newblock \href{https://dx.doi.org/10.1073/pnas.1619152114}{Proceedings of the
  National Academy of Sciences {\bf 114}, 7555--7560}~(2017).

\bibitem{Cao2019}
Yudong Cao, Jonathan Romero, Jonathan~P. Olson, Matthias Degroote, Peter~D.
  Johnson, M{\'{a}}ria Kieferov{\'{a}}, Ian~D. Kivlichan, Tim Menke, Borja
  Peropadre, Nicolas P.~D. Sawaya, Sukin Sim, Libor Veis, and Al{\'{a}}n
  Aspuru-Guzik.
\newblock ``Quantum chemistry in the age of quantum computing''.
\newblock \href{https://dx.doi.org/10.1021/acs.chemrev.8b00803}{Chemical
  Reviews {\bf 119}, 10856--10915}~(2019).

\bibitem{McArdle2020}
Sam McArdle, Suguru Endo, Al{\'{a}}n Aspuru-Guzik, Simon~C. Benjamin, and Xiao
  Yuan.
\newblock ``Quantum computational chemistry''.
\newblock \href{https://dx.doi.org/10.1103/revmodphys.92.015003}{Reviews of
  Modern Physics {\bf 92}, 015003}~(2020).

\bibitem{Farhi2014}
Edward Farhi, Jeffrey Goldstone, and Sam Gutmann.
\newblock ``A quantum approximate optimization algorithm''~(2014).
\newblock  \href{http://arxiv.org/abs/1411.4028}{arXiv:1411.4028}.

\bibitem{Biamonte2017}
Jacob Biamonte, Peter Wittek, Nicola Pancotti, Patrick Rebentrost, Nathan
  Wiebe, and Seth Lloyd.
\newblock ``Quantum machine learning''.
\newblock \href{https://dx.doi.org/10.1038/nature23474}{Nature {\bf 549},
  195--202}~(2017).

\bibitem{Montanaro2016}
Ashley Montanaro.
\newblock ``Quantum algorithms: an overview''.
\newblock \href{https://dx.doi.org/10.1038/npjqi.2015.23}{npj Quantum
  Information{\bf 2}}~(2016).

\bibitem{Peruzzo2014}
Alberto Peruzzo, Jarrod McClean, Peter Shadbolt, Man-Hong Yung, Xiao-Qi Zhou,
  Peter~J. Love, Al{\'{a}}n Aspuru-Guzik, and Jeremy~L. O'Brien.
\newblock ``A variational eigenvalue solver on a photonic quantum processor''.
\newblock \href{https://dx.doi.org/10.1038/ncomms5213}{Nature
  Communications{\bf 5}}~(2014).

\bibitem{McClean2016}
Jarrod~R McClean, Jonathan Romero, Ryan Babbush, and Al{\'{a}}n Aspuru-Guzik.
\newblock ``The theory of variational hybrid quantum-classical algorithms''.
\newblock \href{https://dx.doi.org/10.1088/1367-2630/18/2/023023}{New Journal
  of Physics {\bf 18}, 023023}~(2016).

\bibitem{Cerezo2021}
M.~Cerezo, Andrew Arrasmith, Ryan Babbush, Simon~C. Benjamin, Suguru Endo,
  Keisuke Fujii, Jarrod~R. McClean, Kosuke Mitarai, Xiao Yuan, Lukasz Cincio,
  and Patrick~J. Coles.
\newblock ``Variational quantum algorithms''.
\newblock \href{https://dx.doi.org/10.1038/s42254-021-00348-9}{Nature Reviews
  Physics {\bf 3}, 625--644}~(2021).

\bibitem{Tilly2022}
Jules Tilly, Hongxiang Chen, Shuxiang Cao, Dario Picozzi, Kanav Setia, Ying Li,
  Edward Grant, Leonard Wossnig, Ivan Rungger, George~H. Booth, and Jonathan
  Tennyson.
\newblock ``The variational quantum eigensolver: A review of methods and best
  practices''.
\newblock \href{https://dx.doi.org/10.1016/j.physrep.2022.08.003}{Physics
  Reports {\bf 986}, 1--128}~(2022).

\bibitem{Fedorov2022}
Dmitry~A. Fedorov, Bo~Peng, Niranjan Govind, and Yuri Alexeev.
\newblock ``{VQE} method: a short survey and recent developments''.
\newblock \href{https://dx.doi.org/10.1186/s41313-021-00032-6}{Materials
  Theory{\bf 6}}~(2022).

\bibitem{Nishimori98}
Tadashi Kadowaki and Hidetoshi Nishimori.
\newblock ``Quantum annealing in the transverse ising model''.
\newblock \href{https://dx.doi.org/10.1103/PhysRevE.58.5355}{Phys. Rev. E {\bf
  58}, 5355--5363}~(1998).

\bibitem{Farhi2000}
Edward Farhi, Jeffrey Goldstone, Sam Gutmann, and Michael Sipser.
\newblock ``Quantum computation by adiabatic evolution''~(2000).
\newblock
  \href{http://arxiv.org/abs/quant-ph/0001106}{arXiv:quant-ph/0001106}.

\bibitem{Childs2001}
Andrew~M. Childs, Edward Farhi, and John Preskill.
\newblock ``Robustness of adiabatic quantum computation''.
\newblock \href{https://dx.doi.org/10.1103/physreva.65.012322}{Physical Review
  A {\bf 65}, 012322}~(2001).

\bibitem{AspuruGuzik2005}
{Al\'{a}n Aspuru-Guzik and Anthony D. Dutoi and Peter J. Love and Martin
  Head-Gordon}.
\newblock ``Simulated quantum computation of molecular energies''.
\newblock \href{https://dx.doi.org/10.1126/science.1113479}{Science {\bf 309},
  1704--1707}~(2005).

\bibitem{Albash2018}
Tameem Albash and Daniel~A. Lidar.
\newblock ``Adiabatic quantum computation''.
\newblock \href{https://dx.doi.org/10.1103/revmodphys.90.015002}{Reviews of
  Modern Physics {\bf 90}, 015002}~(2018).

\bibitem{Lee2022}
Seunghoon Lee, Joonho Lee, Huanchen Zhai, Yu~Tong, Alexander~M. Dalzell,
  Ashutosh Kumar, Phillip Helms, Johnnie Gray, Zhi-Hao Cui, Wenyuan Liu,
  Michael Kastoryano, Ryan Babbush, John Preskill, David~R. Reichman, Earl~T.
  Campbell, Edward~F. Valeev, Lin Lin, and Garnet Kin-Lic Chan.
\newblock ``Is there evidence for exponential quantum advantage in quantum
  chemistry?''~(2022).
\newblock  \href{http://arxiv.org/abs/2208.02199}{arXiv:2208.02199}.

\bibitem{Boykin2002}
P.~Oscar Boykin, Tal Mor, Vwani Roychowdhury, Farrokh Vatan, and Rutger Vrijen.
\newblock ``Algorithmic cooling and scalable {NMR} quantum computers''.
\newblock \href{https://dx.doi.org/10.1073/pnas.241641898}{Proceedings of the
  National Academy of Sciences {\bf 99}, 3388--3393}~(2002).

\bibitem{Kaplan2017}
David~B. Kaplan, Natalie Klco, and Alessandro Roggero.
\newblock ``Ground states via spectral combing on a quantum computer''~(2017).
\newblock  \href{http://arxiv.org/abs/1709.08250}{arXiv:1709.08250}.

\bibitem{Wang2017}
Hefeng Wang.
\newblock ``Quantum algorithm for preparing the ground state of a system via
  resonance transition''.
\newblock \href{https://dx.doi.org/10.1038/s41598-017-16396-0}{Scientific
  Reports{\bf 7}}~(2017).

\bibitem{Feng2022}
Jia-Jin Feng, Biao Wu, and Frank Wilczek.
\newblock ``Quantum computing by coherent cooling''.
\newblock \href{https://dx.doi.org/10.1103/physreva.105.052601}{Physical Review
  A {\bf 105}, 052601}~(2022).

\bibitem{Polla2021}
Stefano Polla, Yaroslav Herasymenko, and Thomas~E. O{\textquotesingle}Brien.
\newblock ``Quantum digital cooling''.
\newblock \href{https://dx.doi.org/10.1103/physreva.104.012414}{Physical Review
  A {\bf 104}, 012414}~(2021).

\bibitem{Zaletel2021}
Michael~P. Zaletel, Adam Kaufman, Dan~M. Stamper-Kurn, and Norman~Y. Yao.
\newblock ``Preparation of low entropy correlated many-body states via
  conformal cooling quenches''.
\newblock \href{https://dx.doi.org/10.1103/physrevlett.126.103401}{Physical
  Review Letters {\bf 126}, 103401}~(2021).

\bibitem{Metcalf2020}
Mekena Metcalf, Jonathan~E. Moussa, Wibe~A. de~Jong, and Mohan Sarovar.
\newblock ``Engineered thermalization and cooling of quantum many-body
  systems''.
\newblock \href{https://dx.doi.org/10.1103/physrevresearch.2.023214}{Physical
  Review Research {\bf 2}, 023214}~(2020).

\bibitem{RodriguezThesis}
David Rodr\'{i}guez~P\'{e}rez.
\newblock ``Quantum error mitigation and autonomous correction using
  dissipative engineering and coupling techniques''.
\newblock Ph.D. thesis, Colorado School of Mines~(2021).
\newblock  url:~\url{https://repository.mines.edu/handle/11124/14291}.

\bibitem{Debye1926}
P.~Debye.
\newblock ``Einige bemerkungen zur magnetisierung bei tiefer temperatur''.
\newblock \href{https://dx.doi.org/10.1002/andp.19263862517}{Annalen der Physik
  {\bf 386}, 1154--1160}~(1926).

\bibitem{Giauque1927}
W.~F. Giauque.
\newblock ``A thermodynamic treatment of certain magnetic effects. a proposed
  method of producing temperatures considerably below 1\textdegree absolute''.
\newblock \href{https://dx.doi.org/10.1021/ja01407a003}{Journal of the American
  Chemical Society {\bf 49}, 1864--1870}~(1927).

\bibitem{Cao2021}
Haishan Cao.
\newblock ``Refrigeration below 1 kelvin''.
\newblock \href{https://dx.doi.org/10.1007/s10909-021-02606-7}{Journal of Low
  Temperature Physics {\bf 204}, 175--205}~(2021).

\bibitem{Jones2020}
A.~T. Jones, C.~P. Scheller, J.~R. Prance, Y.~B. Kalyoncu, D.~M. Zumb\"{u}hl,
  and R.~P. Haley.
\newblock ``{Progress in Cooling Nanoelectronic Devices to Ultra-Low
  Temperatures}''.
\newblock \href{https://dx.doi.org/10.1007/s10909-020-02472-9}{Journal of Low
  Temperature Physics {\bf 201}, 772--802}~(2020).

\bibitem{Suppl}
See Supplemental Material for details.

\bibitem{Kibble1976}
T~W~B Kibble.
\newblock ``Topology of cosmic domains and strings''.
\newblock \href{https://dx.doi.org/10.1088/0305-4470/9/8/029}{Journal of
  Physics A: Mathematical and General {\bf 9}, 1387}~(1976).

\bibitem{Zurek1985}
W.~H. Zurek.
\newblock ``Cosmological experiments in superfluid helium?''.
\newblock \href{https://dx.doi.org/10.1038/317505a0}{Nature {\bf 317},
  505--508}~(1985).

\bibitem{Griffin2012}
S.~M. Griffin, M.~Lilienblum, K.~T. Delaney, Y.~Kumagai, M.~Fiebig, and N.~A.
  Spaldin.
\newblock ``Scaling behavior and beyond equilibrium in the hexagonal
  manganites''.
\newblock \href{https://dx.doi.org/10.1103/PhysRevX.2.041022}{Phys. Rev. X {\bf
  2}, 041022}~(2012).

\bibitem{Dutta2016}
Anirban Dutta, Armin Rahmani, and Adolfo del Campo.
\newblock ``Anti-kibble-zurek behavior in crossing the quantum critical point
  of a thermally isolated system driven by a noisy control field''.
\newblock \href{https://dx.doi.org/10.1103/PhysRevLett.117.080402}{Phys. Rev.
  Lett. {\bf 117}, 080402}~(2016).

\bibitem{Dziarmaga2005}
Jacek Dziarmaga.
\newblock ``Dynamics of a quantum phase transition: Exact solution of the
  quantum ising model''.
\newblock \href{https://dx.doi.org/10.1103/PhysRevLett.95.245701}{Phys. Rev.
  Lett. {\bf 95}, 245701}~(2005).

\bibitem{Zurek2005}
Wojciech~H. Zurek, Uwe Dorner, and Peter Zoller.
\newblock ``Dynamics of a quantum phase transition''.
\newblock \href{https://dx.doi.org/10.1103/PhysRevLett.95.105701}{Phys. Rev.
  Lett. {\bf 95}, 105701}~(2005).

\bibitem{Polkovnikov2005}
Anatoli Polkovnikov.
\newblock ``Universal adiabatic dynamics in the vicinity of a quantum critical
  point''.
\newblock \href{https://dx.doi.org/10.1103/PhysRevB.72.161201}{Phys. Rev. B
  {\bf 72}, 161201}~(2005).

\bibitem{Barankov2008}
Roman Barankov and Anatoli Polkovnikov.
\newblock ``Optimal nonlinear passage through a quantum critical point''.
\newblock \href{https://dx.doi.org/10.1103/PhysRevLett.101.076801}{Phys. Rev.
  Lett. {\bf 101}, 076801}~(2008).

\bibitem{Nishimori21}
Yuki Susa and Hidetoshi Nishimori.
\newblock ``Variational optimization of the quantum annealing schedule for the
  lechner-hauke-zoller scheme''.
\newblock \href{https://dx.doi.org/10.1103/PhysRevA.103.022619}{Phys. Rev. A
  {\bf 103}, 022619}~(2021).

\bibitem{Matthies2024}
Anne Matthies, Mark Rudner, Achim Rosch, and Erez Berg.
\newblock ``{Data for "Programmable adiabatic demagnetization for systems with
  trivial and topological excitations" [Data set]}''.
\newblock \href{https://dx.doi.org/10.5281/ZENODO.10805566}{Zenodo}~(2024).

\end{thebibliography}


\clearpage
\onecolumn
\appendix

\section{Cooling rate in the perturbative regime}
Here, we derive the cooling rate in the small coupling $g$ limit, i.e., we calculate the derivative of the expectation value of the system Hamiltonian $H_s$, 
\begin{equation}
	\langle \dot{H_s}\rangle= i \langle \psi(t)| [H_c,H_s] |\psi(t) \rangle, \label{appE:start}
\end{equation}
where  $H_c=g(t)\sum_{i=1}^N \hat{A}_i^s \sigma_i^y $ describes the coupling between the system and the bath. The local degrees of freedom of the system, $\hat{A}_i^s$, are coupled to $N$ bath spins described by the spin operators $\mathbf{\sigma}_i$.
In the interaction picture, we can use perturbation theory to find the wave function $|\psi(t)\rangle=U_I(t)|\psi(0)\rangle$ of the system and bath for small $g(t)$ by expanding the time evolution operator $U_I(t)$ up to first order in $g(t)$:
\begin{equation}
	U^I(t,0)\approx \mathbb{1}-i \int_{0}^t H_c(t') dt'. \label{appE:UI}
\end{equation}

With this expansion, we can simplify Eq.~(\ref{appE:start}) up to 2nd order in $g(t)$ to
\begin{align}
	\langle \dot{H_s}\rangle&\approx
	g(t) \int_{0}^t dt' \langle \psi(0)|[[\hat{A}_i^s(t) \sigma_i^y(t),H_s],H_c(t')] |\psi(0) \rangle  \nonumber \\
	&=  -ig(t)\int_{0}^t dt' g(t')\sum_{j}(\langle \hat{A}_j^s(t')\dot{\hat{A}}_j^s(t)\rangle \langle \sigma_j^y(t') \sigma_j^y(t)\rangle 
 + \text{c.c.},\label{appE:Hsdot}
\end{align}
where we used that $i\langle \psi(0)| [H_c,H_s] |\psi (0) \rangle =0$ and that spin and bath are initially not entangled.

In the following, we will calculate the cooling rate per site $\langle \dot{H}\rangle /N$ with the system contribution $\Sigma_S(t,t')=\langle \hat{A}_i^s(t')\dot{\hat{A}}_i^s(t)\rangle$ and the bath contribution $\Sigma_B(t,t')=\langle \sigma_y(t') \sigma_y(t)\rangle$, where $N$ is the number of system spins.

We assume that the system is thermalizing and we can use the fluctation-dissipation theorem to calculate the correlation function 
\begin{align}
	\Sigma_s(\omega)&= 2i \omega (1+n_B(\omega)) \chi_s^{\prime\prime}(\omega), \label{appE:sysresponse}
\end{align}
where $\chi_s^{\prime\prime}(\omega)$ is the imaginary part of the local $\hat{A}^s_i$ susceptibility of our system and  $n_B(\omega)$ is the Bose function.

To further evaluate Eq.~(\ref{appE:Hsdot}), we need to calculate for the bath spins $\sigma^y(t)=e^{i\int H_b (t'){dt'}}\sigma_i^y e^{-i\int H_b (t'){dt'}}$, where $H_b=-B(t)\sum_i\sigma_i^z$ is the bath Hamiltonian and $\theta_B(t)=-\int_0^t B(t'){dt'}$.
Direct evaluation of the exponentials yields
\begin{align}
	\sigma^y_i(t)
	&=\sin(2\theta_B(t))\sigma_i^x + \cos(2\theta_B(t))\sigma_i^y.
\end{align}

Using that initially the bath spins are polarized in the $+\hat z$ direction, we obtain
\begin{align}
	\langle \sigma_j^y(t') \sigma_j^y(t)\rangle
	&=e^{2i(\theta_B(t')-\theta_B(t))} \label{appE:bathspins}
\end{align}

Inserting Eq.~(\ref{appE:sysresponse}) and (\ref{appE:bathspins})
into Eq.(\ref{appE:Hsdot}), we find the cooling rate per site as
\begin{align}
	\frac{\langle \dot{H_s}\rangle}{N}	
&	=2  \int  {d}\omega [ \omega (1+n_B(\omega))\chi^{\prime\prime}_y(\omega) 
 g(t)  \int_{0}^t {dt'} g(t') e^{2i(\theta_B(t')-\theta_B(t))} e^{+i\omega (t-t')}]+c.c. \label{appE:Hdotf}
\end{align}

Integrating the second line of this expression over time allows to calculate how much energy is extracted from a mode with frequency $\omega$ during a cooling cycle: 
\begin{align}
\Delta_c(T,\omega)=\label{delc} 
\int_{0}^T \! dt \, g(t) \int_{0}^t \! dt' g(t') e^{2i(\theta_B(t')-\theta_B(t))} e^{+i\omega (t-t')}+c.c.
\end{align}
With this definition, the total energy per site transferred from the system to the bath is given by the frequency integral \begin{align}
 \frac{\Delta E_{\rm tot}}{N}=2 \int  {d}\omega \, \omega (1+n_B(\omega))\chi^{\prime\prime}_s(\omega)\,
\Delta_c(T,\omega).    
\end{align}
In our protocol, we sweep the magnetic field linearly down, $B(t)=B_i-\Gamma_B t$, such that $\partial_t B(t)=\Gamma_B$.

In the adiabatic limit of a small sweep rate $\Gamma_B$ one can evaluate Eq.~\eqref{delc} analytically using a stationary phase approximation for the two time integrals. The main contribution comes for $t$ and $t'$ close to $t^*$ with $\omega=- 2 B(t^*)$, i.e., when the magnetic field is at resonance with the frequency $\omega$ (the factor $2$ arises as the energy $2 B$ is needed for a spin-flip of a bath spin). Using a Taylor expansion of second order around that point we obtain.
\begin{align}
\frac{\Delta_c(T,\omega)}{g(t^*)^2}&\approx
\int_{0}^T {dt}\int_{0}^t {dt'} e^{-i  \left( (t-t^*)^2 -(t'-t^*)^2 \right)\Gamma_B}+c.c.\nonumber \\
&=
\int_{0}^T {dt}\int_{0}^{T} d t' e^{-i  \left( (t-t^*)^2 -(t'-t^*)^2 \right)\Gamma_B}\nonumber \\
&\approx \frac{\pi}{\Gamma_B} \quad \text{for } \ \Gamma_B \to 0,\ t^*<T, \label{eq:adia}
\end{align}
where we used, that $t^*$ and $T$ scale with $1/\Gamma_B$.
Finally, we arrive at
\begin{equation}
   \Delta_c(T,\omega)\approx g(t^*)^2 \frac{\pi}{\Gamma_B}(1-\theta(\omega+2B(T))).
    \label{appE:bathapprox}
\end{equation} 
\begin{figure}[t]
\centering
		\includegraphics[width=0.49\columnwidth]{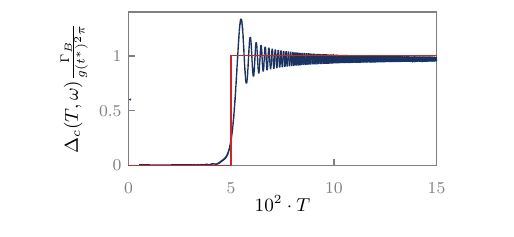}
	\caption{\label{Appfig:pertubative} The bath absorption rate $\Delta_c(T,\omega)$ (blue), defined in Eq.~\eqref{delc}, as a function of sweep duration $T$ in the perturbative analysis for $B_i=1$, $B_f=0.001$ and $\omega=-1$ compared to the approximation of Eq.~\eqref{appE:bathapprox}.}
\end{figure}
\begin{figure}[t]
\centering
	\includegraphics[width=0.49\columnwidth]{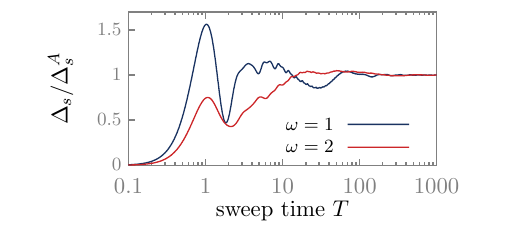}
	\caption{\label{fig:pertubative} Plot of $\Delta_s(T,\omega)=\Delta_c(T,\omega)-\Delta_c(T,-\omega)$ as a function of $T=1/B_i$ for  $\omega=1$ (blue) and  $\omega=2$ (red).
		$\Delta_s$ is normalized to the value computed from the adiabatic approximation, $\Delta_s^A=\frac{ \pi g(t^*)^2}{\Gamma_B}$, see Eq.~\eqref{eq:adia}.
		We use a sweeping protocol with $B(t)=5(1-t/T)$. The system-bath coupling $g(t)$ increases linearly for $t<T/4$ from $0$ to some finite value $g$, remains constant for $T/4<T<3 T/4$, and drops linear to $0$ for $t>3 T/4$.}  
\end{figure}
Figure~\ref{Appfig:pertubative} compares the approximate expression in Eq.~(\ref{appE:bathapprox}) with the result of direct numerical integration, and shows that the Heaviside function is indeed a good approximation.
Finally, we can calculate $\Sigma(t, \omega)$ by taking the derivative of $A(t,\omega)$
\begin{align}
\Sigma(t, \omega)= \partial_\tau A(t,\omega)=2 \pi g(t^*)^2 \delta(\omega+2B(t))
\label{appE:batht}
\end{align}
In the end, we arrive at cooling rate per site
\begin{align}
	 \frac{\langle \dot{H_s}\rangle}{N}
&	\approx 4 \pi g(t)^2 \int {d}\omega \delta(\omega+2B(t)) \omega (1+n_B(\omega)) \chi_s^{\prime\prime}(\omega),	\label{chiH}
\end{align}
where we inserted the bath contribution in the stationary phase approximation, Eq.~(\ref{appE:batht}), into Eq.~(\ref{appE:Hdotf}) for the cooling rate. The equation, equivalent to Eq.~(3) of the main text, shows that the energy of a mode with frequency $\omega$ is extracted most efficiently when the resonant condition $\omega=-2 B(t)$ is met. This is rigorously valid only in the adiabatic limit. 

Instead of using the adiabatic approximation, we can also investigate  directly $\Delta_c(T,\omega)$ numerically to explore how efficient one can extract energy using more rapid sweeps of the magnetic field. In  Fig.~\ref{fig:pertubative} we plot
$\Delta_s(T,\omega)=\Delta_c(T,\omega)-\Delta_c(T,-\omega)$ as a function of the duration $T$ of a cooling cycle. This combination takes into account that both cooling (removal of $\omega$) and heating (addition of $\omega$) processes occur via a mode with frequency $\omega$. The relative weight of processes with $\omega$ and $-\omega$  depends via the $n_B(\omega)$ term in Eq.~\eqref{appE:Hdotf} on the temperature of the system; for the plot we consider the case where the temperature is much larger than $\omega$, $n_B(\omega)\approx \frac{T}{\omega} \approx -n_B(-\omega)$. Therefore $\Delta_c(T,\omega)$ and $\Delta_c(T,-\omega)$ contribute with the same weight in this case.

In the figure, we normalize $\Delta_s$  to the value obtained via an adiabatic approximation, $\Delta_s^A=\frac{ \pi g(t^*)^2}{B}$, see Eq.~\eqref{eq:adia}. 
For a very short cooling-cycle duration $T$, energy extraction is not efficient and strongly suppressed. For very large $T$ one recover the result from the adiabatic approximation. For intermediate $T$ the extracted energy depends on details of the sweeping protocol and the frequency $\omega$ but in general an efficient energy extraction is already possible for remarkably short $T$.

\section{Cooling rate for small densities of bath spins}

In our study, we mainly focus on the case where the numbers of system and bath spins are equal.
However, on small quantum devices one may also run the same algorithm for a much smaller number of bath spins. In this  section, we discuss how the cooling rate depends on the ratio $f=\frac{N_\text{bath}}{N_\text{system}}$ of bath and system spins in the small-$f$ limit.

In this limit, the bath spins are far apart and one has to investigate whether the bottleneck for cooling
is given by the bath spins or by the transport of energy to the location of the spins.
In the small-$g$ limit analyzed above, the bottleneck is the coupling to the bath spins and thus the cooling rate is simply proportional to $f$,  $\Gamma_c \propto f$.

For finite $g$ and in one-dimensional systems, the bottleneck for cooling for small $f$ is the transport of energy to the site where the bath spin is located over distances of order $1/f$. For example, if the heat transport is diffusive (assuming that the system is non-integrable with a mean-free path small compared to $1/f$), then the time scale needed to cool spins far from the bath site scales with $1/f^2$. Thus, we expect that $\Gamma_c \propto f^2$ in the limit $f \to 0$ at finite $g$. A closely related result can be found in Ref. \cite{Zaletel2021}, where the cooling power of a single link between a system-chain and a bath-chain has been investigated.

In higher dimension, the distance of equally spaced bath spins is proportional to $(1/f)^{1/d}$.
In dimensions $d>2$,  the bottleneck is not the transport of energy to the bath site (which occurs on the time-scale $1/f^{2/d}$) but simply the fact that there are only very few cooling sites, resulting in $\Gamma_c\sim f$.
In $d=2$, one obtains from the same argument also $\Gamma_c\sim f$ (with possible logarithmic corrections) independent of whether the bottleneck for cooling is given by the diffusion of energy or by the low density of cooling sites.

\section{Non-integrable models}
\begin{figure*}[]
	\centering{
	\includegraphics[width=\columnwidth]{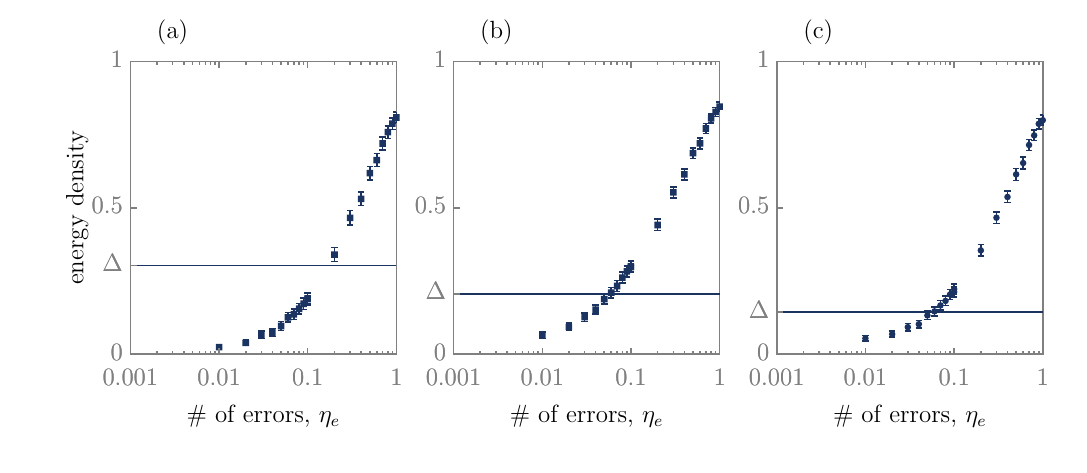}}
	\caption{\label{fig:energy noise appendix} $(E-E_0)/E_0$ as a function of the average number of errors $\eta_e$ for a Hamiltonian with integrability-breaking perturbations: (a) with longitudinal field ($J=0.4$, $h_x=0.5$, $h_z=0.8$, $J_x=0$), 
 and with exchange interaction $J_x$ (b) with parameters ($J=1$, $h_x=0.5$, $h_z=0$, $J_x=0.2$) and (c) with parameters  ($J=0.4$, $h_x=0.7$, $h_z=0$, $J_x=0.1$). The solid line indicates the lowest excitation gap $\Delta$.  (Parameters: $N_s=8$, $N_{\tau}=101$, $N_c=100$, $N_{init}=1000$, $T=6$, $B_i=5$, $B_f=0.7$, $g_0=0.5$).}
\end{figure*}
In Fig.~3 of the main text, we focus on the comparison between the paramagnetic and ferromagnetic phase using two `dual' versions of the transverse field Ising model. This model is, however, special as it is (in the absence of coupling to the bath spins and to noise) integrable and can be solved exactly using fermionization.

To confirm that we obtain a similar performance of our cooling protocol, we show in this section also results for the final-state energy density as function of noise rate for three non-integrable models.

In Fig.~\ref{fig:energy noise appendix} a, we use a longitudinal field to break integrability, in Fig.~\ref{fig:energy noise appendix} b and c we consider the model
\begin{align}
	H_s=-\sum_{i} \left( J s_i^z s_{i+1}^z + J_x s_i^x s_{i+1}^x \right)- \sum_{i} h_x s_i^x.     \label{eq:Ising non int}
\end{align}
Compared to the model of Eq. (4) in the main text, we have added an extra integrability breaking interaction of strength $J_x$.

A comparison of Fig.~\ref{fig:energy noise appendix} to Fig.~3 in the main text shows that qualitatively similar cooling efficiencies are obtained. Furthermore, Fig.~6 in the main text also shows that the non-integrable model shown in panel b has a similar energy distribution compared to the integrable cases, a and c.

\end{document}